# Random Telegraph Noise based True Random Number Generator for Fully Integrated Systems

Gilson Wirth, *Senior Member, IEEE*, Pedro A B Alves, and Roberto da Silva

*Abstract* — Generating streams of true random numbers is a critical component of many embedded systems. The design of fully integrated, area and power efficient True Random Number Generators is a challenge. We propose a fully integrated, lightweight implementation, that uses the random telegraph noise (RTN) of standard MOSFET as entropy source. It is not analog-intensive, and without traditional post-processing algorithms, the generated random bit sequence passes the National Institute of Standards and Technology (NIST) tests.

## I. INTRODUCTION

Random number generators (RNGs) are a crucial component of many embedded systems. They are used for data encryption, secure communication, among other complex processes [1-4]. It also has a wide range of applications in machine learning and stochastic logic [5, 6].

RNGs can be classified as True RNGs (TRNGs) and Pseudo RNGs (PRNGs). TRNGs are based on physical entropy sources and are inherently much more secure, because future or past values cannot be inferred from present values. PRNGs are deterministic systems generating bit sequences that seem to be random, but their entire trajectory can be predicted once the seed or an intermediate state is known [3].

Unfortunately, the much higher level of security in TRNGs is obtained at the cost of larger area and power consumption [1, 3]. The demand for ubiquitous security down to tightly constrained systems – such as IoT – has motivated extensive investigation of low-power and inexpensive True Random Number Generators (TRNGs) [7].

In this work, we propose a cheap, light, small and reliable TRNG that uses as entropy source the random telegraph noise (RTN) produced by standard MOSFETs, readily available in any integrated circuit. The proposed approach is not analog-intensive, and to avoid costly and energy hungry post-processing, simple bit truncation is used to harvest the entropy from the RTN source.

The circuitry for entropy generation, entropy extraction, and its utilization (e.g., for encryption) may be fully implemented in a single integrated circuit. Standalone entropy generation and extraction may expose the system to physical attacks - as for instance bus micro probing - due to the physical separation of the different modules [1, 3]. Standalone entropy generation and extraction may also lead to increased area and energy consumption.

The proposed approach passes the NIST tests [8].

## II. RTN AS ENTROPY SOURCE FOR TRNG

In MOSFETs, charge trapping and de-trapping at localized states (charge traps) at the interface or in the gate dielectric generates discrete fluctuations in the device conductance. These fluctuations are called random telegraph noise (RTN). Figure 1 depicts the charge tapping mechanism. In an electrically biased MOSFET, this capture and emission of charge carriers by electrically active defects (charge traps) produces discrete fluctuations of the drain current $I_D$. $I_D$ alternates between a high current state (when the trap is vacant) and a low current state (when the trap is occupied), as seen in Fig. 2. The average time in the high current state (state 1) is $\tau_C$, while $\tau_E$ is the average time spent in the low current state (state 0) [9].

If we take the voltage at the drain of a MOSFET in a simple common source stage, the device conductance fluctuations will be seen as drain voltage fluctuations. These drain voltage fluctuations may be used as the entropy source (labeled "Original RTN") of a fully integrated true random number generator, as discussed below.

The time between capture and emission events is an exponential random variable [14]. The times are independent from previous events and the number of events that occur in a fixed time period follows a Poisson distribution. The randomness in times between charge capture and/or emission events can be converted into random digits in a few different ways, in a similar approach as used in TRNGs based on single photon detectors and Geiger counters [1].

Geiger counters and single photon detectors are among the earliest used entropy sources for TRNG, but they have limitations, such as limited counting capabilities. Another limitation to most single photon detectors and Geiger counters is the time they need to recover after a detection, known as dead time [1]. Regarding Geiger counters, the need for a radioactive source is an issue. Natural activity rarely produces enough

This work was supported in part by CNPq, CAPES and FAPERGS.
Corresponding author: Gilson Wirth (e-mail: gilson.wirth@ufrgs.br).
Gilson Wirth is with the Electrical Engineering Department, UFRGS, Porto Alegre, RS, Brazil.
Pedro A B Alves was with the Electrical Engineering Department, UFRGS, Porto Alegre, RS, Brazil.
Roberto da Silva is with the Physics Institute, UFRGS, Porto Alegre, RS, Brazil.
Color versions of one or more of the figures in this article are available online at http://ieeexplore.ieee.org

particles to cause more than a few counts per second [1]. Regarding photon detectors, they are not easily integrated into embedded systems. Using RTN as an entropy source solves these issues.

In the approach here proposed, the random time between capture and emission events is used to generate the random number. A fast clock – fast if compared to the time between capture and emission events – is used to increase a counter. If a capture or emission event happens, the counter is read and reset. The value read from the counter is used to produce the random number. Figure 3 gives a graphical description of the method. In the implementation shown, the counter is read and reset at both falling and rising edges of the digitized RTN signal, i.e., at both charge capture and emission events. Alternatively, the counter could be read and reset only at falling or rising edges. I.e., only at charge capture or only at charge emission event.

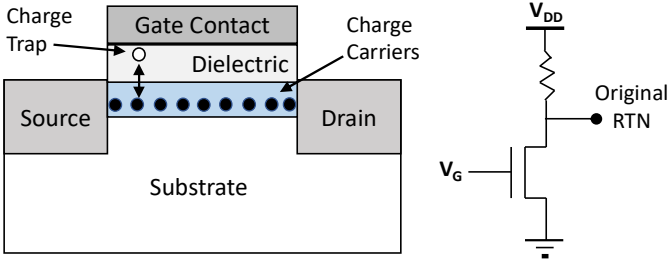

**Fig. 1.** Left side: Capture and emission of mobile charge carrier by electrically active defects – charge traps – results in RTN. Right side: a simple common source stage, where the RTN appears as drain voltage fluctuations.

The time between capture and emission events is expected to be an exponentially distributed random variable. The goal is to obtain a uniformly distributed random variable. By applying truncation, the least significant bits of the digitized time between capture and/or emission events (counter reading) may be a uniformly distributed random variable, avoiding the need for costly post-processing. Using the least significant bits to avoid post-processing was already implemented in works that used entropy sources different from RTN [1]. Post-processing could be implemented, with the drawback of silicon area and energy consumption, which may be an issue for IoT applications, but not for larger embedded systems. The post-processing stage may convert the raw bit sequence into a good quality output as close as possible to a uniform bit distribution and can also include tasks like checking if the generator is working properly (randomness testing). There are different post-processing algorithms to convert most of the randomness available in the exponential distribution into a uniform bit sequence (requiring additional hardware and energy consumption) [1]. Using the least significant bits of the digitized time is a simple and efficient way, and achieves the goal of implementing a TRNG, as shown in the next section.

If larger bit rates are needed, the generated random numbers could be used as a random seed for pseudorandom number generators (PRNGs). Starting from a random seed, PRNGs may yield a longer output random string of good randomness.

Examples include generating a random seed for a Non-Linear Feedback Shift Register (NLFSR) [1, 8, 10].

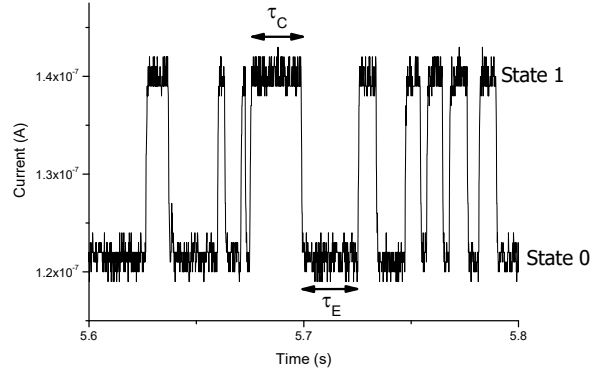

**Fig. 2.** Random Telegraph Noise (RTN): charge capture and emission by a trap leads to drain current switching between discrete states, labeled "*State 0*" and "*State 1*". $\tau_C$ and $\tau_E$ are the capture and emission time constants, respectively.

### III. IMPLEMENTATION OF THE RTN BASED TRNG

We present a simple and efficient solution for the hardware implementation of the TRNG. The RTN produced by a MOSFET is used as the entropy source - making it compatible with standard semiconductor technology - and it can be readily implemented by basic circuit modules. Other MOS technology compatible devices – such as ReRAM – may also be used as the RTN source [11].

We use a digital counter that increases its output value by 1 when it receives a pulse at its input and can be reset to restart the count from 0. A digital clock is used to increment the counter. The digital counter is incremented at a rate much faster than the rate of charge capture and emission events, i.e., the digital clock is much faster than the rate at which the RTN changes state. The value of the counter at the moment of the charge capture or emission event is read and used to produce the random number. The two most significant bits are discarded.

We show the block diagram for the simple circuit implementation explored in this work in Fig. 4. The RTN may be taken at the drain of a MOSFET, using a simple common source stage, as shown in the right side of Fig. 1. It is labeled "Original RTN" in the block diagram of Fig. 4. The first step is to digitize the Original RTN signal. It is compared to a reference voltage ($V_{REF}$), yielding the digitized RTN at the output of the comparator. $V_{REF}$ may be obtained, for instance, from a low-pass filter. Passing the RTN through a low-pass filter may yield the average (DC) value of the RTN, which is an appropriate reference to digitize the RTN signal.

A transition of the RTN signal – i.e., a charge capture or emission event – triggers the reading of the counter. The counter output is read to a memory element. The memory element may be simple flip-flops or latches. The next step is the truncation of the value read from the counter. In this implementation, the two most significant bits are discarded. The remaining bits form the random bit stream. Such an empirical process seems to be an essential mechanism to generate a random number that satisfies the criteria required in the literature.

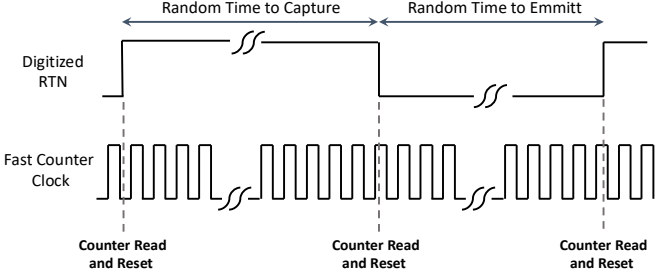

**Fig. 3.** A fast clock is used to increment a counter. Whenever a transition at the digitized RTN happens (capture or emission event), the counter is read and reset, generating one random number.

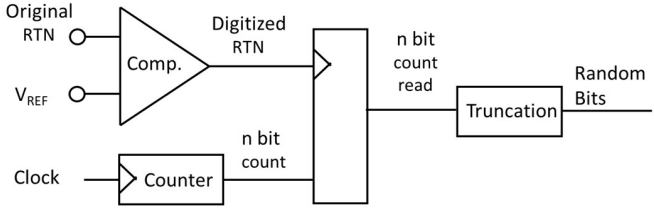

**Fig. 4.** Block diagram for the circuit implementation used in this work.

### IV. VALIDATION

The implementation described above was simulated. In the actual circuit implementation, the RTN comes from a MOSFET. In the simulation of the circuit, one needs to simulate the RTN. To simulate a RTN with entropy similar to the actual RTN of a MOSFET, we use true random numbers from the Australian National University Quantum Random Number Generator (ANU QRNG) [12].

It is crucial to notice the need to use true random numbers from the ANU QRNG to evaluate the RTN transition probability (charge capture and emission events) since if we use pseudo-random numbers from the computer's processor, the generated bit stream is also pseudo-random and the NIST test fails, as discussed below.

In the simulation, $\Delta t$ is the time step of the discrete time simulation, which is assumed to be 1 (one) arbitrary time unit. For a 0→1 transition, the transition probability is $P_{0\rightarrow 1} = \Delta t/\tau_E$. For a 1→0 transition, the transition probability is $P_{1\rightarrow 0} = \Delta t/\tau_C$ [9]. The RTN time series is simulated as depicted in Fig. 5.

For the simulation here performed, we attribute to both $\tau_C$ and $\tau_E$ the value of 2500, in arbitrary time units. The period of the fast clock that increments the counter is 1 arbitrary time unit.

The transition probability is compared to a true random number between 0 and 1, obtained from the ANU QRNG. If a transition – trap capture or emission event – happens, the counter is read and reset. The two most significant bits read are discarded, and the remaining bits are used to form the output random bit stream. The number of random bits generated was 111072. The string of 111072 bits is then submitted to the NIST test suite [8]. As can be seen in Table I, it passes all tests, validating the approach proposed here. It is important to note that the Maurer's Universal Statistical Test was not performed [13]. The Maurer's Universal Statistical Test requires extremely long sequence lengths, much longer than the generated string of 111072 bits [8]. One of the reasons is the realization that asymptotic approximations are used in determining the limiting distribution. Due to the computational cost, generating such extremely long bit streams in our simulation is prohibitive.

If instead of true random numbers obtained from the ANU QRNG, random numbers obtained from the microprocessor's native pseudo random number generator are used for evaluating RTN transition probability in the numerical simulation, the NIST test fails. This is not a surprise, since the random numbers generated by microprocessor's native pseudo random number generator also failed the NIST test. This emphasizes the need for TRNGs that can be integrated in digital systems, including IoT, microprocessors and embedded systems in general.

*For each time step $\Delta t$ during simulation*
    *Get random number between 0 and 1 from ANU QRNG*
    *if (RTN_state = 0)*
      *if ($P_{0\rightarrow 1}$ > random number)*
        *RTN_state = 1*
    *else if (RTN_state = 1)*
      *if ($P_{1\rightarrow 0}$ > random number)*
        *RTN_state = 0*

**Fig. 5.** Pseudocode for the generation of the RTN time series. The capture and emission of charge carriers is a Markov process that depends only on the current state. A capture or emission event generates a RTN state change. The randomness in times between RTN state changes is used as entropy source.

TABLE I
NIST TEST SUITE RANDOMNESS TEST RESULTS

| Test Name | P-Value | Result |
|---|---|---|
| Frequency Test (Monobit) | 0.0797 | Pass |
| Frequency Test within a Block | 0.1226 | Pass |
| Run Test | 0.9547 | Pass |
| Longest Run of Ones in a Block | 0.9703 | Pass |
| Binary Matrix Rank Test | 0.1384 | Pass |
| Discrete Fourier Transform | 0.0941 | Pass |
| Non-Overlapping Template Matching | 0.7826 | Pass |
| Overlapping Template Matching | 0.3795 | Pass |
| Linear Complexity Test | 0.9454 | Pass |
| Serial Test | 0.8663 | Pass |
| Approximate Entropy Test | 0.6811 | Pass |
| Cumulative Sums (Forward) Test | 0.1544 | Pass |
| Cumulative Sums (Reverse) Test | 0.0364 | Pass |
| Random Excursions Test | 0.7189 | Pass |
| Random Excursions Variant Test | 0.9535 | Pass |

## V. Conclusion

In this letter, we proposed a hardware implementation of a power and area efficient True Random Number Generator that uses the Random Telegraph Noise of standard MOSFETs as an entropy source. It may be implemented in a single integrated circuit. Without traditional post-processing algorithms, the generated bit stream passes the National Institute of Standards and Technology (NIST) randomness tests. The next step of our work is to design and fabricate an integrated circuit with the proposed implementation.